# INTEGRATION OF GEOELECTRIC AND GEOCHEMICAL DATA USING SELF-ORGANIZING MAPS (SOM) TO CHARACTERIZE A LANDFILL

Camila Juliao[1], Johan Diaz[1], Yosmely Bermúdez[1], Milagrosa Aldana[2],
[1]Coordinación de Ingeniería Geofísica. Universidad Simón Bolívar. Caracas, Venezuela.
[2]Departamento de Ciencias de la Tierra. Universidad Simón Bolívar. Caracas, Venezuela.
14-10540@usb.ve, 15-10418@usb.ve, 13-10143@usb.ve, maldana@usb.ve

**ABSTRACT**

Leachates from garbage dumps can significantly compromise their surrounding area. Even if the distance between these and the populated areas could be considerable, the risk of affecting the aquifers for public use is imminent in most cases. For this reason, the delimitation and monitoring of the leachate plume are of significant importance. Geoelectric data (resistivity and IP), and surface methane measurements, are integrated and classified using an unsupervised Neural Network to identify possible risk zones in areas surrounding a landfill. The Neural Network used is a Kohonen type, which generates; as a result, Self-Organizing Classification Maps or SOM (Self-Organizing Map). Two graphic outputs were obtained from the training performed in which groups of neurons that presented a similar behaviour were selected. Contour maps corresponding to the location of these groups and the individual variables were generated to compare the classification obtained and the different anomalies associated with each of these variables. Two of the groups resulting from the classification are related to typical values of liquids percolated in the landfill for the parameters evaluated individually. In this way, a precise delimitation of the affected areas in the studied landfill was obtained, integrating the input variables via SOMs. The location of the study area is not detailed for confidentiality reasons.

*Keywords*: Neural Networks, Self-organizing Maps, Landfill, Resistivity, Chargeability, Methane.

**INTRODUCTION**

Clustering is one of the most useful tools and is commonly used in data analysis, particularly with geophysical data. The size and dimensionality of existing geospatial databases emphasize the need for efficient and robust clustering algorithms. Self-organizing maps (SOMs) are one of these data analysis and integration tools. SOMs are a type of artificial neural network, which are trained using unsupervised competitive learning to produce a low-dimensional, discretized representation of the set of training samples. SOMs are also called "feature maps" since they preserve the main characteristics of the input data (Miljković, 2017). Thus, SOMs allow different parameters to be integrated and analyzed together, through a dimensional reduction of the data. The output is a two-dimensional graph that organizes the samples using unsupervised learning (Sakao and Neramballi, 2020). Self-organizing maps are known for their clustering, visualization and classification capabilities (e.g. Penn (2005); Dossary et al. (2016); Carter-McAuslan and Farquharson (2020)). Strecker and Uden (2002), for example, use SOMs to classify post-stack seismic attributes in a 3D data volume from Lafourche Parish, South Louisiana. Analysis of the self-organized maps allowed them to interpret stratigraphic characters that were masked in the conventional amplitude volume. Kohler et al. (2010) use this type of neural network to identify temporal patterns in seismic vibration records. The objective is to characterize, locally, an active volcano in Indonesia. SOM analysis allows them to visualize and group characteristics such as frequency, wavenumber, polarization, and spectral analysis. As a result, they are able to identify the wave fields that affect the quality of the Love wave estimates.

Given its usefulness in pattern recognition when several variables are analyzed, the objective of this study is to obtain SOM maps to characterize an area in which a landfill is located and aquifers for urban use are present. Aquifers for public use require constant monitoring as a consequence of the exponential growth in solid waste production that significantly affects them. Currently, there are non-invasive geophysical methods that allow this task to be carried out. However, these can lead to ambiguities in the interpretation of the results if the data derived from different methods are analyzed individually. The lack of accuracy in interpretation may occur if the lithologies present in the area give physical responses similar to those of the leachates, for example equivalent resistive responses (Abdulrahman et al., 2016). In the present study, resistivity measurements, induced polarization (IP) and methane measurements, previously obtained in the area, will be used as input data to generate the SOM and characterize the landfill. It is expected to obtain graphic outputs that group the data more similarly and help identify possible areas associated with pollution plumes.

**THEORETICAL ASPECTS**

Neurobiological studies indicate that different inputs, coming from the different senses of the human being, that is, sight, hearing, smell, touch, movement, etc., are assigned to particular areas of the cerebral cortex in an orderly way. This can be thought of as a brain topographic map. This type of cortical maps has, in principle, two important properties. On the one hand, at each stage of data processing, each piece of incoming information is preserved in its own context or neighborhood. Additionally, neurons that process closely related information are kept together to interact through short synaptic connections (Morales, 2018). Based on this behavior, Kohonen (1982) proposes an Artificial Neural Network model capable of effectively mapping similar patterns (i.e. input data vectors that, due to their similarity, are closer) to contiguous locations in the output space. The aim is, therefore, to build artificial topographic maps that learn through self-organization, inspired by neurobiology (Kohonen, 2001). According to Vesanto (2000), the SOM is a bilayer neural network. It consists of an input layer and an output layer. The components of a SOM are the nodes or neurons, which have an associated weight vector, of the same dimension as that of the input vectors, and a position on a topographic map. The common configuration of neurons in a regular two-dimensional space is a grid, usually hexagonal or rectangular. Thus, the model is composed of a vector of weights, equivalent to the synaptic connections of a real neuron, an input formed by N neurons to be selected and an output scalar of the neurons that process the information and forms the feature map. The weights on each connection or link correspond to the intensity of the equivalent synapse. These weights are determined during the training or learning process and are usually started randomly (Torrecilla et al., 2009). Once the weights are initialized, a vector is selected from the input data set, randomly or sequentially, and the neuron or node closest to the input vector is determined, according to its weights. The winning node will correspond to the Best Matching Unit or BMU. To determine the BMU, different distance measures can be used (Euclidean, Cosine, Minkowski, Inner Product, among others). Various licenses and some computing platform applications, such as Matlab's "nctool", use a Euclidean distance. The weights of the BMU, as well as those of those nodes close to it in the SOM grid, are modified to bring them closer to the input vector. The change in neighboring nodes depends on a neighborhood function (NF), so that the weights of the neurons closest to the BMU are modified more than those of the neurons further away. The above process is iterated for the input data, decreasing the NF with the number of iterations. Different NFs can be used and the convergence speed of the SOM can depend on the selection (Vesanto, 2000; Ota et al., 2011; Kohonen, 2013). SOMs have been implemented using a function that assigns a value of 1 to the neurons closest to the BMU and zero to those furthest away; Gaussian functions, used by applications such as MATLAB's "nctool"; the negative of the second derivative of a Gaussian function, which is strongly attractive at a small scale, somewhat repulsive at a medium scale and slightly attractive at a large scale, and even asymmetric neighborhood functions. According to this process, successive and different inputs will cause corrections in different subsets or areas of the initial grid configuration. Kohonen (2013) points out that, as the entire spatial neighborhood around the BMU in the grid is modified at the same time and this modification propagates gradually over it, according to the NF, the degree of local ordering increases with the number of iterations. With this process, the SOM associates the output neurons with groups or patterns from the training data set. Summarizing the process, each input will be associated with a weight vector that it modifies, causing the network to learn, and the outputs are a function of this modification. Using a neighborhood function to preserve the topological properties of the input space makes self-organizing maps different from other unsupervised neural networks.

**METHODOLOGY**

To carry out the study, vectors of the type (Resistivity, Chargeability, Methane), measured in 107 locations in the study area, were used as input data for the SOM. The data were previously measured and provided for this work. For confidentiality reasons, neither the study area nor the work group that assigned them is specified. The data were organized in a matrix of dimension 107x3. The MATLAB "nctool" module was used to obtain the SOMs. This module allows to manually select the appropriate number of neurons to classify the variables in the data set. In this case, the process of selecting the number of neurons to generate the final SOM maps was carried out by iterating the training of the self-organizing map with different network sizes. Networks of 9x9, 8x8, 7x7, 6x6, 5x5, 4x4 and 3x3 neurons were tested, so the final number of these corresponds to the multiplication of the respective dimensions. An hexagonal topology was selected and 200 epochs were used for training in all cases. The criterion for selecting the appropriate dimension for this particular data set was based on the observation of the number of neurons that were activated during training, the number of samples contained in each of them, and the distances between the neighborhoods of network weights. This was evaluated from the two graphical outputs of the SOM: the first shows the arrangement of the data in the neurons and the second the distance between them, as exemplified in Figures 1 and 2, respectively, for a 4x4 network. In the distance map, the black color represents the furthest distance between the neurons and, therefore, between the data contained in each group associated with a center. The yellow color corresponds to the closest distance.

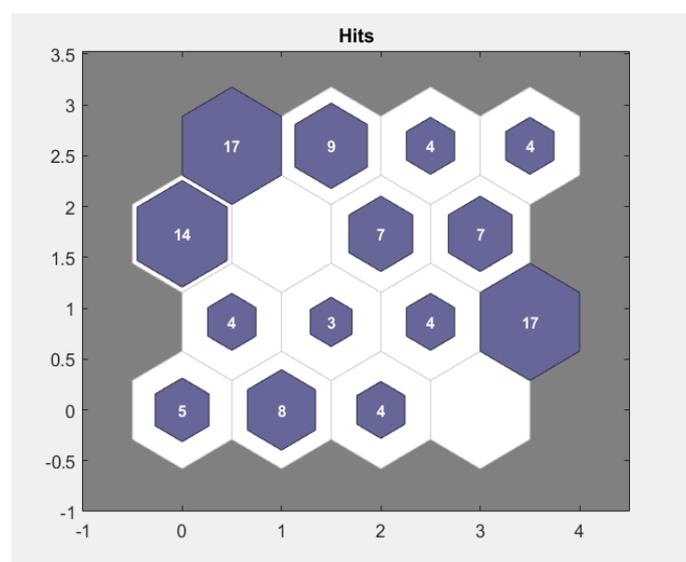

**Figure 1.** *Example of the arrangement of the input data of this study, in the neurons of the self-organized map of a 4x4 network.*

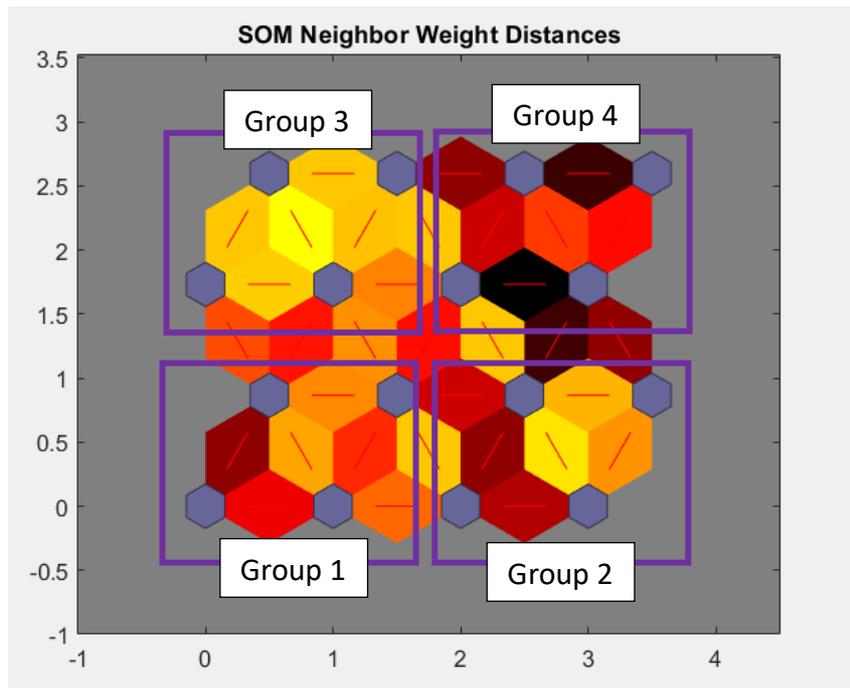

**Figure 2.** *Example of a map of distances between the weights of the SOM neighborhood, labeling the interpreted groups, for a 4x4 network.*

Once the appropriate network size is selected, the data contained in each of the neurons is extracted with an additional routine coded in MATLAB. By integrating this information, groups of neurons that presented similar behavior in terms of proximity were selected to obtain a contour map that showed the location of these groups, according to the samples contained in each one. In addition, in order to make a comparison between the classification obtained and the different anomalies associated with resistivity, induced polarization and surface methane concentration, additional maps were generated for each variable; these maps are shown in Figures 3 to 5, respectively. It is important to highlight that the black points observed in these figures and that are surrounded by the respective contours, correspond to the acquisition points in the field.

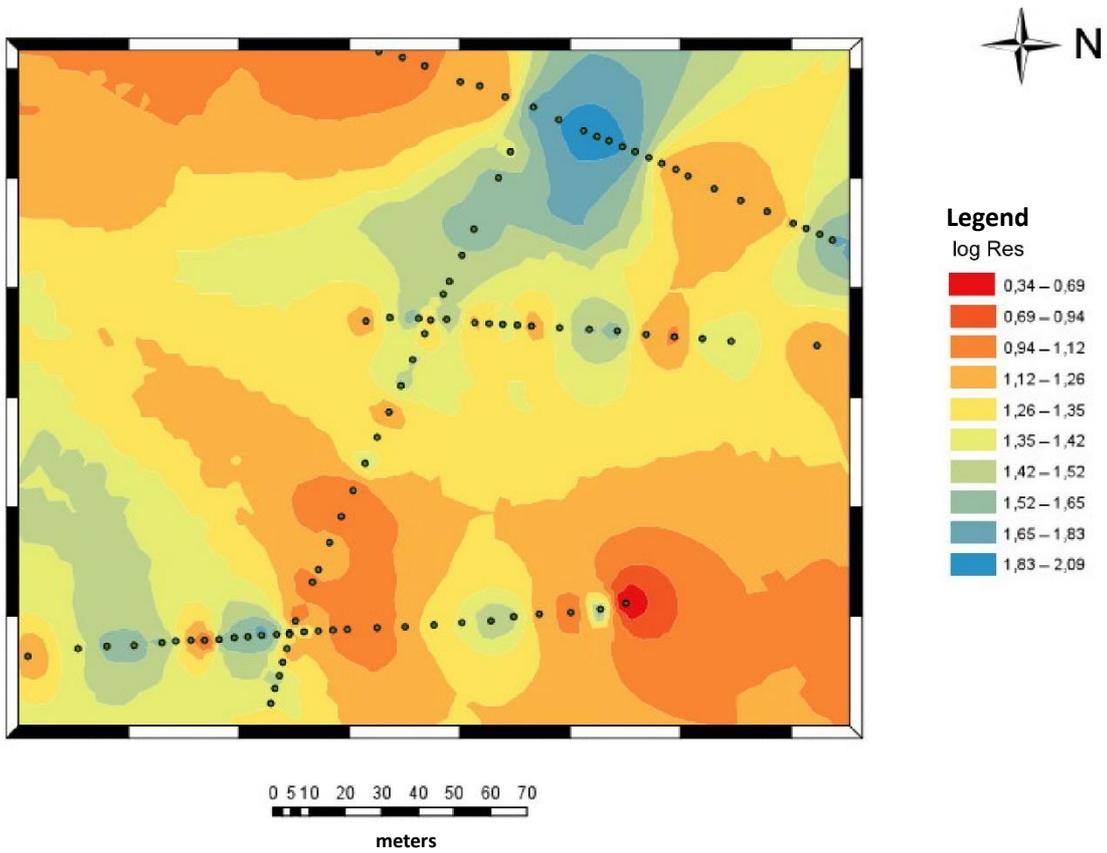

**Figure 3.** *Map of the logarithm of resistivity.*

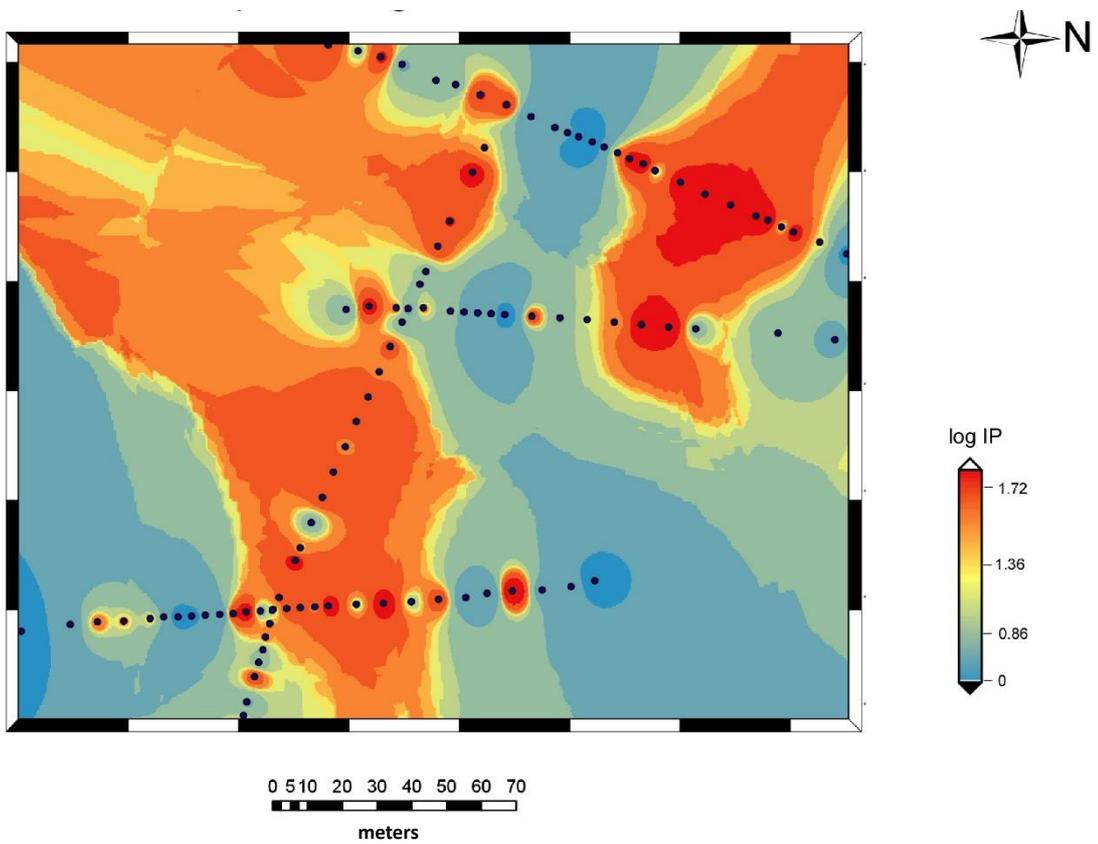

**Figure 4.** *Map of the logarithm of induced polarization.*

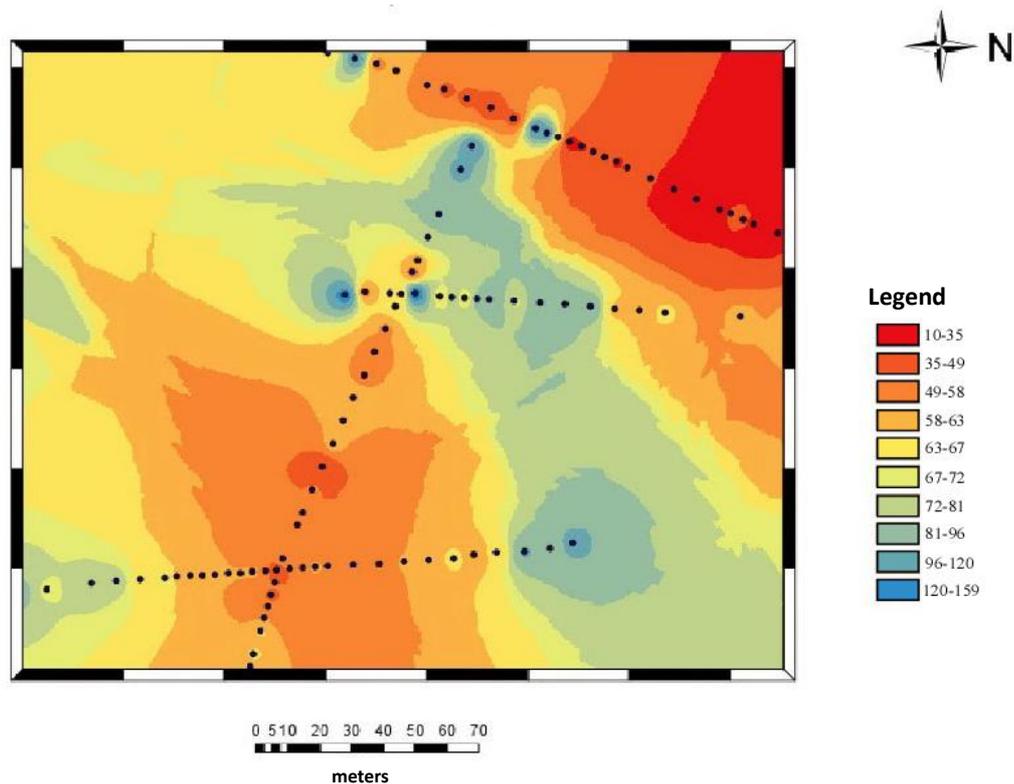

**Figure 5.** *Map of the logarithm of methane concentration, in ppm, on the surface*

## RESULTS

Figure 6 shows the results obtained for the 9x9, 6x6 and 3x3 network sizes. For network sizes between 9x9 and 5x5, several deactivated neurons were observed when grouping the 107 data entered and no considerable groups were obtained in the distance map between them. In the case of the 3x3 network, although there were no deactivated neurons, it was not possible to differentiate clusters. The data layout for each neuron in the 4x4 network is shown in Figure 1. As can be seen when comparing with Figure 6, this network is the one with the fewest inactive neurons (only two deactivated neurons). In addition, 4 groups or clusters of neurons can be identified, using the graphic output of the SOM that shows the distance between neighboring neurons (Figure 2). Thus, according to the previously established criteria, the 4x4 network was selected for data analysis.

The arrangement of the samples in groups of neurons, according to this selection, can be seen in the contour map of Figure 7. For this last graph, groups 1 and 2 correspond to the area with the colors red to orange and Groups 3 and 4 to the area of blue tones.

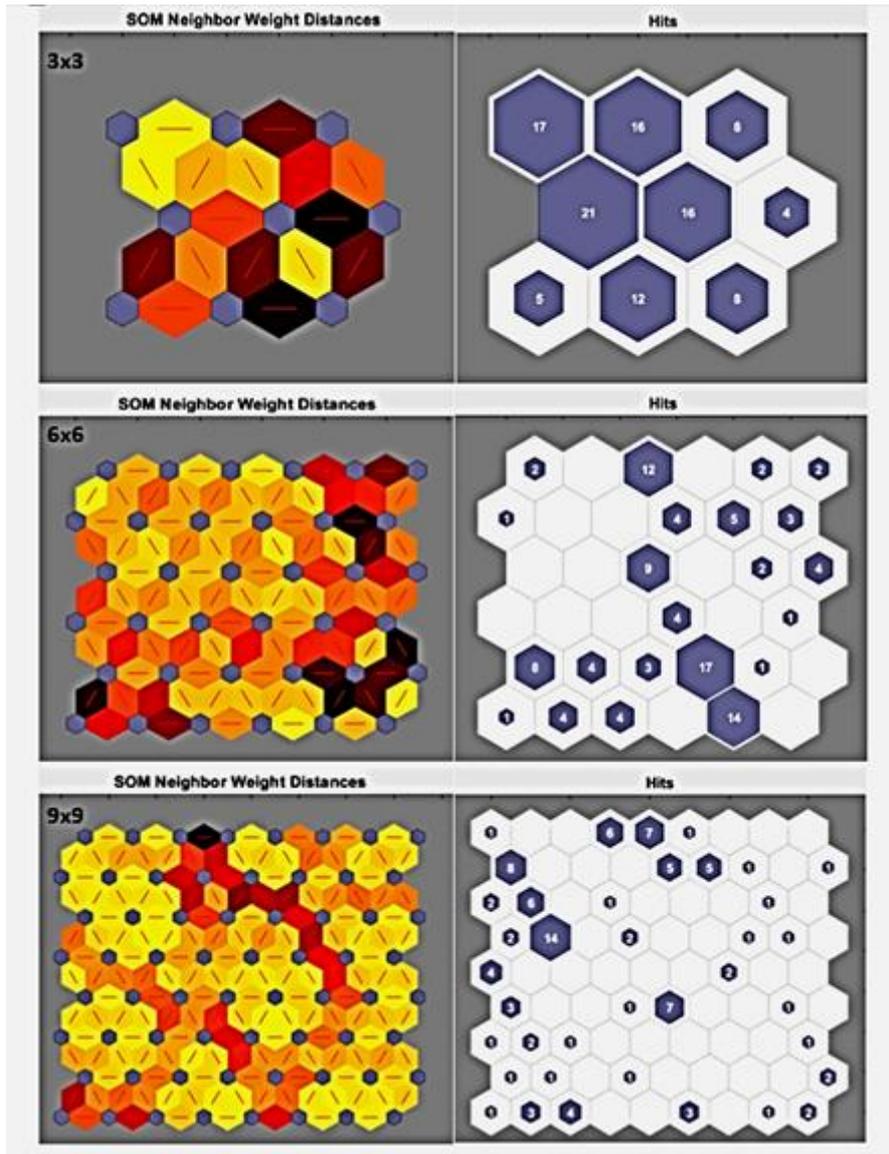

**Figure 6.** *Maps of distances between the SOM neighborhood weights and the disposition of the input data in the neurons, for the 3x3, 6x6 and 9x9 networks, respectively.*

As observed in Figure 7 and as specified in Table 1, groups 1 and 2 of neurons coincide with low values of the logarithm of resistivity, in a range of 0.34 to 0.94, and high values of the logarithm of IP and the logarithm of methane, in a range between 1.36 and 1.72, and between 81 to 159, respectively. This can be appreciated by comparing the SOM maps with the contour maps of the individual parameters (Figures 3 to 5) and verifying this with the data contained in the neurons. Thus, the data in groups 1 and 2 correspond to low values of resistivity and high values of chargeability and methane concentration, which is a typical response of percolated liquids (Abdulrahman et al., 2016). This suggests the existence of a leachate plume in the studied landfill, whose aquifers are located at a depth of approximately 25 m, and which is associated with the areas where these groups of neurons are located.

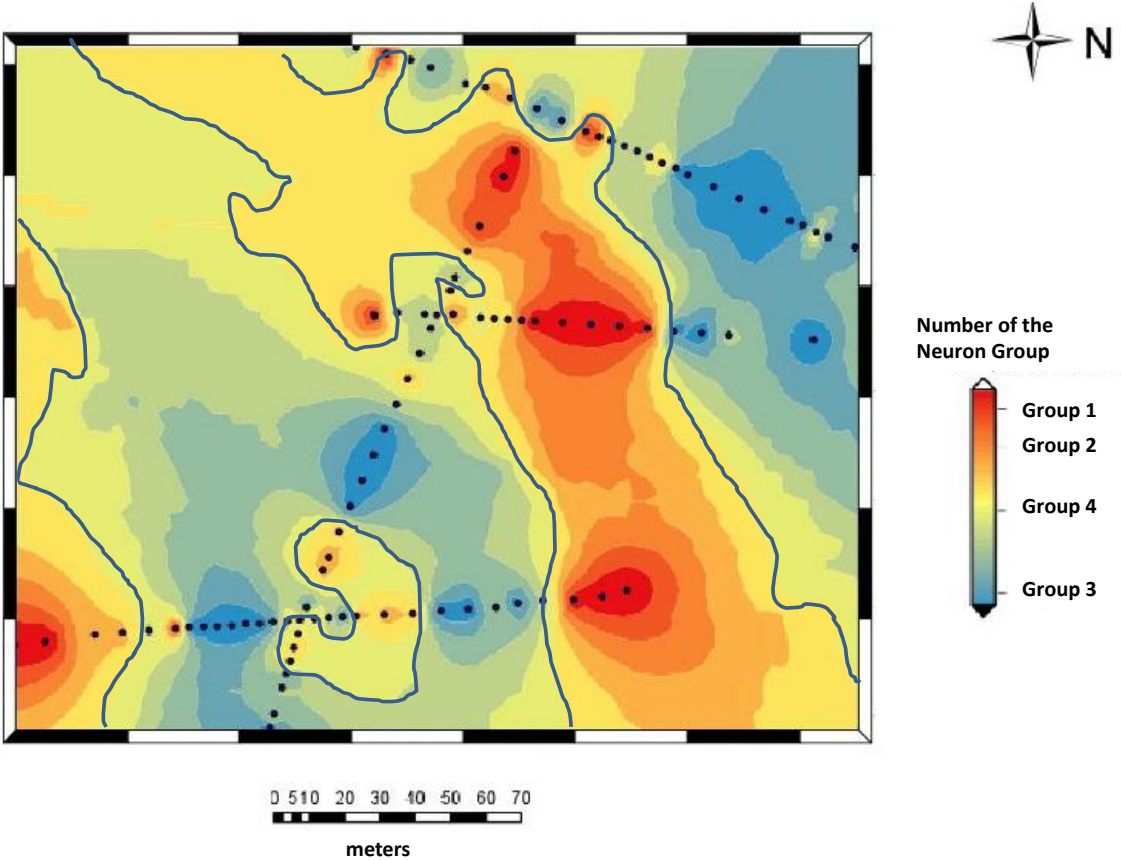

**Figure 7.** *Contour map corresponding to the 4 groups identified from the SOM analysis. Groups 1 and 2, associated with the possible presence of leachate, are enclosed with a blue line. The ranges of logarithm of resistivity, logarithm of IP and methane concentration (ppm) associated with each group are presented in table 1.*

**Table 1.** *Ranges of values of the parameters Logarithm of Resistivity, Logarithm of IP and methane concentration (ppm), associated with each of the 4 groups identified in the SOM.*

| Group | Log(Res) | Log(IP) | Methane (ppm) |
|---|---|---|---|
| 1 | 0.34-0.69 | 1.36-1.54 | 81-105 |
| 2 | 0.69-0.94 | 1.57-1.72 | 110-159 |
| 3 | 1.42-2.09 | 0.91-1.35 | 10-49 |
| 4 | 1.12-1.42 | 0.14-0.89 | 49-72 |

The map in Figure 7 delimits the possible leachate plume, enclosing, with a blue line, groups 1 and 2 resulting from the SOM. It is possible to observe an elongated, well-defined morphology, with an approximate NW-SE direction, which would correspond to a leachate plume that we will designate as the main plume. Towards the S and SW, two other areas, possibly associated with leachates, are observed. These last two zones could be, in principle, the result of one of the following two reasons. A possible explanation is that they originate from the migration of leachates, from the main plume towards the south-southwest of the area, and accumulate in that zone, due to the characteristics of the terrain. However, they could also be the result of the presence of another source of contaminants towards the SW, which gives rise to a different plume. However, there is no additional information about the area that would allow us to corroborate whether these areas respond to the existence of another plume or to the migration and partial accumulation of leachates from the main identified plume.

It is important to note that, if the input variables (resistivity, induced polarization or surface methane concentration) are analyzed separately, certain areas of the landfill could be delimited as areas affected by percolated liquids. However, some lithologies may have physical responses similar to that of leachates, depending on the parameter measured, which could lead to an interpretation with a certain degree of error when analyzing these parameters individually.

## CONCLUSIONS

In the present work, we characterize the area of a landfill, integrating three types of geophysical data (resistivity, chargeability and methane gas concentration measurements), using an unsupervised network, Kohonen type, to generate self-organizing maps (SOM). The SOMs effectively integrated and classified these data, allowing delimiting possible areas affected by the landfill and associated with the leachate migration plume. The recognition of patterns with these maps, which integrate the response of different variables, is an advantage of using SOMs to classify and recognize trends in this type of studies.

## ACKNOWLEDGMENTS

The authors thank Drs. María Jácome and Vincenzo Costanzo-Álvarez for the support provided.